\documentclass[a4paper]{jpconf}
\usepackage{graphicx}
\def \d {{\rm d}}
\begin{document}
\title{Cylindrically and toroidally symmetric solutions with a cosmological constant}

\author{Ji\v{r}\'{\i} Podolsk\'{y}$^1$ and Jerry Griffiths$^2$}

\address{$^1$ Institute of Theoretical Physics, Charles University in Prague,\\  V Hole\v{s}ovi\v{c}k\'{a}ch 2, 180 00 Prague 8, Czech Republic}
\address{$^2$ Department of Mathematical Sciences, Loughborough University,\\ Loughborough, LE11~3TU, UK}
\ead{podolsky@mbox.troja.mff.cuni.cz}

\begin{abstract}
Cylindrical-like coordinates for constant-curvature 3-spaces are introduced and discussed. This helps to clarify the geometrical properties, the coordinate ranges and the meaning of free parameters in the static vacuum solution of Linet and Tian. In particular, when the cosmological constant is positive, the spacetimes have toroidal symmetry. One of the two curvature singularities can be removed by matching the Linet--Tian vacuum solution across a toroidal surface to a corresponding region of the dust-filled Einstein static universe. Some other properties and limiting cases of these space-times are also described, together with their generalisation to higher dimensions.
\end{abstract}

\section{Introduction}
The class of static cylindrically symmetric vacuum solutions, which was found by Levi-Civita in 1919 [1], can be written in the form
\begin{equation}
\d s^2 =-\rho^{4\sigma/\Sigma}\d t^2 + \rho^{-4\sigma(1-2\sigma)/\Sigma}\d z^2 +C^{2}\rho^{2(1-2\sigma)/\Sigma} \d\phi^2 +\d\rho^2,
\label{LeviCivita}
\end{equation}
where ${\Sigma=1-2\sigma+4\sigma^2}$. The parameter ${\sigma\in(0,{1\over4})}$ may be interpreted as the mass per unit length of the source located along the axis ${\rho=0}$, while $C$ is the conicity parameter (see [2] for more details). When ${\sigma=0}$, Minkowski space in cylindrical coordinates is recovered.

In 1986, a generalisation of (\ref{LeviCivita}) to include a non-zero cosmological constant~${\Lambda}$ was obtained by Linet~[3] and Tian~[4] (see also [5]) in the form
\begin{equation}
 \d s^2=Q^{2/3}\Big( -P^{-2(1-8\sigma+4\sigma^2)/3\Sigma}\,\d t^2
 +P^{-2(1+4\sigma-8\sigma^2)/3\Sigma}\,\d z^2
 +C^{2}P^{4(1-2\sigma-2\sigma^2)/3\Sigma}\,\d\phi^2 \Big)
 +\d\rho^2,
\label{LinetTianmetric}
\end{equation}
 where $\rho$ is a proper radial distance from the axis and
 \begin{equation}
 Q(\rho)={1\over\sqrt{3\Lambda}}\sin\Big(\sqrt{3\Lambda}\,\rho\Big), \qquad
 P(\rho)={2\over\sqrt{3\Lambda}}\tan\bigg({\sqrt{3\Lambda}\over2}\,\rho\bigg).
 \label{PQdefs}
 \end{equation}
Both ${\Lambda>0}$ and ${\Lambda<0}$ are admitted (in the latter case the trigonometric functions are replaced by hyperbolic ones and $\Lambda$ by ${|\Lambda|}$). The metric (\ref{LinetTianmetric}), (\ref{PQdefs}) locally approaches the Levi-Civita solution (\ref{LeviCivita}) either as ${\Lambda\to0}$ or near the axis as ${\rho\to0}$ because ${Q\approx\rho }$ and  ${P\approx\rho }$ in these limits.

\section{Constant-curvature 3-spaces in cylindrical coordinates}
In order to understand the global geometrical properties of the Linet--Tian solution, it seems to be important first to rewrite the {\it maximally symmetric} 3-spaces in cylindrical-like coordinates. Such constant-curvature spaces are usually written in the spherical form
\begin{equation}
\d s^2=R^2\,\Big(\,\frac{\d r^2}{1-k\,r^2}+r^2(\d \theta^2+\sin^2\theta\,\d \phi^2)\Big),
\label{constcurvspher}
\end{equation}
where ${k=0, +1, -1}$ corresponds to the geometries of ${E^3,\, S^3,\, H^3}$, respectively (as is well-known from the FLRW cosmology). Performing the transformation
\begin{equation}
\hat\rho=r\,\sin \theta, \qquad \hat z,\, \tan\hat z,\, \tanh\hat z=\frac{r\,\cos\theta}{\sqrt{1-k\,r^2}},
\label{transf}
\end{equation}
the metric (\ref{constcurvspher}) becomes
\begin{equation}
\d s^2=R^2\Big( (1-k\,\hat\rho^2)^{-1}\d\hat\rho^2 +(1-k\,\hat\rho^2)\,\d\hat z^2  + \hat\rho^2\,\d\phi^2\Big).
\label{constcurvcyl}
\end{equation}
In the case when ${k=0}$, this is the flat space $E^3$ in cylindrical coordinates. Since ${\hat z\in(-\infty,+\infty)}$, \ ${\phi\in[0,2\pi)}$, the surfaces ${\hat\rho=\,}$const. are obviously {\it cylinders} with topology $R^1\times S^1$.

\begin{figure}[ht]
\begin{center}
\includegraphics[scale=0.7]{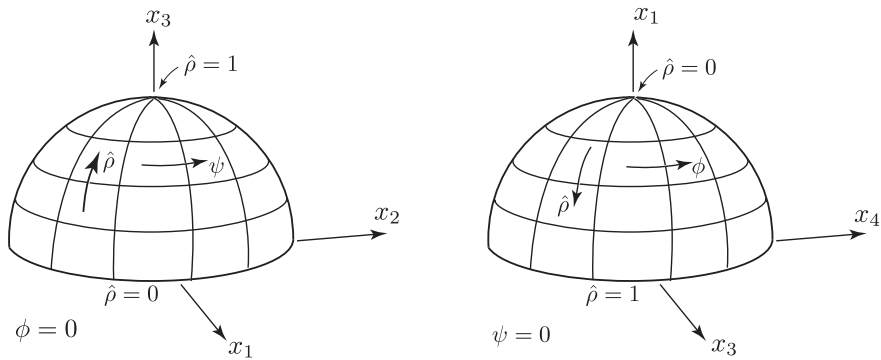}
\end{center}
\caption{\label{f1} Two sections through the three-sphere $S^3$, namely ${\phi=0}$ (left) and ${\psi=0}$ (right). Here $\psi$ and $\phi$ are ``complementary'' angular coordinates in different directions. Thin tori ${\hat\rho=\,}$const. around the axes ${\hat\rho=0}$ and ${\hat\rho=1}$ do not intersect. }
\end{figure}

In the case when ${k=+1}$, the metric of the three-sphere $S^3$ in cylindrical-like coordinates is
\begin{equation}
\d s^2=R^2\left( \frac{\d\hat\rho^2}{1-\hat\rho^2} +(1-\hat\rho^2)\,\d\psi^2  + \hat\rho^2\,\d\phi^2\right),
\label{3spherecyl}
\end{equation}
where ${\psi=\hat z-{\pi\over2}}$. Since the 3-space is bounded in this case, both $\psi$ and $\phi$ are {\it periodic} coordinates with ${\psi,\phi\in[0,2\pi)}$, ${\hat\rho\in[0,1]}$, and the surfaces ${\hat\rho=}$~const.~are {\em tori} with topology ${S^1\times S^1}$. There are thus {\em two} nonintersecting circular axes at ${\,\hat\rho=0\,}$ and ${\,\hat\rho=1}$ around the closed space $S^3$. This can be explicitly seen from the parametrisation of $S^3$ as the 3-surface ${{x_1}^2+{x_2}^2+{x_3}^2+{x_4}^2={R}^2}$ in a flat space ${\d s^2=\d{x_1}^2+\d{x_2}^2+\d{x_3}^2+\d{x_4}^2}$. The metric of (\ref{3spherecyl}) is obtained by
\begin{eqnarray}
&& \hspace*{-13mm} x_1=R\,\sqrt{1-\hat\rho^2}\,\cos\psi, \hspace*{4mm}  x_3=R\,\hat\rho\,\cos\phi,
\hspace*{3mm} \hbox{i.e.}, \hspace*{3mm}
x_1^2+x_2^2=R^2(1-\hat\rho^2),   \hspace*{4mm}  x_3^2+x_4^2=R^2\hat\rho^2,  \nonumber\\
&& \hspace*{-13mm}  x_2=R\,\sqrt{1-\hat\rho^2}\,\sin\psi,  \hspace*{4.5mm}  x_4=R\,\hat\rho\,\sin\phi,
\hspace*{20.8mm} \frac{x_2}{x_1}=\tan\psi,  \hspace*{20.8mm}  \frac{x_4}{x_3}=\tan\phi,
\label{paramet}
\end{eqnarray}
see figure~\ref{f1}.
\vspace{-8.7mm}

\section{Interpreting the Linet--Tian solution with $\Lambda>0$}
In view of the above geometry, it seems necessary to relabel $z$ in (\ref{LinetTianmetric}) as the angular coordinate $\psi$ and to introduce the related conicity parameter $B$. The Linet--Tian vacuum metric becomes
\begin{equation}
 \d s^2=Q^{2/3}\Big( -P^{-2(1-8\sigma+4\sigma^2)/3\Sigma}\,\d t^2 +B^{2}P^{-2(1+4\sigma-8\sigma^2)/3\Sigma}\,\d\psi^2  +C^{2}P^{4(1-2\sigma-2\sigma^2)/3\Sigma}\,\d\phi^2 \Big)
 +\d\rho^2,
\label{LinetTianmetric2}
\end{equation}
where ${Q(\rho), P(\rho)}$ are given by (\ref{PQdefs}) and ${\phi,\psi\in[0,2\pi)}$. Apart from the cosmological constant~$\Lambda$, the metric has three parameters: $B$, $C$ and $\sigma$. There are curvature singularities along the ``axes'' ${\rho=0}$ and ${\rho=\pi/\sqrt{3\Lambda}}$, with the deficit angles ${2\pi(1-C)}$ and ${2\pi(1-B)}$ in the weak-field limit.

Either of these two singularities can be removed and specific values of $B$ and $C$ can be established, for example, by {\it matching} this vacuum solution across a surfaces on which $\rho$ is constant (${\rho=\rho_1}$) to a corresponding {\it toroidal region of the Einstein static universe}. This well-known homogeneous and isotropic dust-filled universe has the spatial geometry of $S^3$ with radius ${R=1/\sqrt\Lambda}$ and thus, using the toroidal coordinates of (\ref{3spherecyl}), can be written in the form
\begin{equation}
\d s^2= -A_1^2\d t^2
 + \frac{B_1^{2}}{\Lambda} \cos^2\Big(\sqrt\Lambda\,(\rho-\rho_{0})\Big)\d \psi^2
 + \frac{C_1^{2}}{\Lambda} \sin^2\Big(\sqrt\Lambda\,(\rho-\rho_{0})\Big)\d \phi^2
 +\d\rho^2.
\label{Einstein}
\end{equation}
For convenience, we applied a simple transformation ${\hat\rho\equiv\sin\Big(\sqrt\Lambda\,(\rho-\rho_{0})\Big)}$ and introduced the free constants $A_1, B_1, C_1$ and $\rho_0$  (see [5]).

Usual matching conditions, namely that the metrics (\ref{LinetTianmetric2}) and (\ref{Einstein}) and their first derivatives are continuous across the surface ${\rho=\rho_1}$, can indeed be consistently satisfied when
 \begin{eqnarray}
 && \hspace*{-16mm}\cos\Big(\sqrt{3\Lambda}\,\rho_1\Big)={1-8\sigma+4\sigma^2\over1-2\sigma+4\sigma^2},\qquad\quad
\tan^2\Big(\sqrt\Lambda\,(\rho_1-\rho_0)\Big) = {4\sigma(1-\sigma)\over1-4\sigma},
\label{match12}\\
 A_1 \!\!\!\!&=&\!\!\!\! Q(\rho_1)^{1/3}\>P(\rho_1)^{-(1-8\sigma+4\sigma^2)/3\Sigma}, \label{match3}\\
 B_1 \!\!\!\!&=&\!\!\!\! B\,\sqrt{\Lambda}\, Q(\rho_1)^{1/3}\>P(\rho_1)^{-(1+4\sigma-8\sigma^2)/3\Sigma} /\cos\Big(\sqrt\Lambda(\rho_1-\rho_{0})\Big), \label{match4} \\
 C_1 \!\!\!\!&=&\!\!\!\! C\,\sqrt{\Lambda}\, Q(\rho_1)^{1/3}\>P(\rho_1)^{2(1-2\sigma-2\sigma^2)/3\Sigma} /\sin\Big(\sqrt\Lambda(\rho_1-\rho_{0})\Big), \label{match5}
 \end{eqnarray}
which uniquely determine ${\rho_1, \rho_0}$ in terms of ${\sigma\in[0,\frac{1}{4}]}$ and relate ${A_1, B_1, C_1}$  to ${B,C}$.
\vspace{-10mm}
\begin{figure}[h]
\includegraphics[scale=0.62]{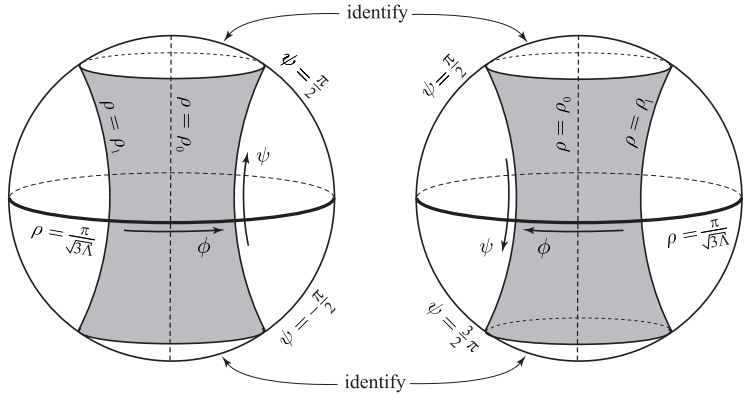}\hspace{1pc}%
\begin{minipage}[b]{13.4pc}\caption{\label{f2} Toroidal region of the Einstein static universe (grey) serves as the dust matter source of the Linet--Tian vacuum solution.}
\end{minipage}
\end{figure}

In the resulting composite spacetime, the curvature singularity at ${\rho=0}$ is removed. As shown in figure~\ref{f2}, it is replaced by the {\it toroidal region} ${\rho\in[\,\rho_0, \rho_1)}$ which is a part of uniform Einstein static space filled with dust --- {\it the matter source} (this is regular at ${\rho_0}$ when ${C_1=1}$). In the {\it external region} ${\rho\in[\,\rho_1,\pi/\sqrt{3\Lambda})}$ there is the {\it Linet--Tian static vacuum solution}. However, there still remains the curvature singularity at ${\rho=\pi/\sqrt{3\Lambda}}$ (which could alternatively be removed by a complementary toroidal dust matter source, keeping the singularity at ${\rho=0}$).

It is now straightforward to calculate the total mass of such a toroidal dust source of density ${\mu=\frac{\Lambda}{4\pi}}$, yielding
${\int_{\rho_0}^{\rho_1}\!\!\int_0^{2\pi}\!\!\int_0^{2\pi}\mu\,\sqrt{g_3}\,\d\rho\,\d\psi\,\d\phi
 =\frac{2\pi B_1C_1}{\sqrt\Lambda}\,\frac{\sigma(1-\sigma)}{(1-4\sigma^2)}}$. The {\it mass per unit length} of the toroid is thus ${\sigma(1-\sigma)/(1-4\sigma^2)}$, i.e., it is determined just by the parameter $\sigma$. This demonstrates that $\sigma$ in the Linet--Tian class of solutions with ${\Lambda\not=0}$ retains its physical meaning known from the Levi-Civita spacetime (for the discussion of possible shell sources see [6]).

Interestingly, the ``no source'' limit ${\sigma=0}$ {\it is not the (anti-)de~Sitter space}, as one would naturally expect! For ${\sigma=0}$, the Linet--Tian metric reduces to
\begin{equation}
 \d s^2 =p^2(-\d t^2+B^{2}\,\d\psi^2) +\frac{4C^2}{3\Lambda}\frac{(1-p^3)}{p}\,\d\phi^2 +\frac{3}{\Lambda}\frac{p}{(1-p^3)}\,\d p^2,
\label{nosource}
\end{equation}
where the coordinate $p\equiv\cos^{2/3}\Big(\frac{\sqrt{3\Lambda}}{2}\,\rho\Big)$ was introduced. This spacetime of algebraic type~D belongs to the Pleba\'nski--Demia\'nski family and also to the Kundt family of solutions. In fact, it is a generalization of the BIII metric [3,7]. Its geometrical properties still need to be investigated.

\section{Extension to higher dimensions}
The Linet--Tian class of solutions described above can be extended to any higher $D$-dimensions. Such toroidally symmetric static vacuum metrics read
\begin{equation}
\d s^2=R(\rho)^\alpha\bigg( -S(\rho)^{2p_0}\,\d t^2  +\sum_{i=1}^{D-2}C_i^{2}\,S(\rho)^{2p_i}\,\d\phi_i^{\,2} \bigg)+\d\rho^2,
\label{HDG}
\end{equation}
where
\begin{equation}
R(\rho)=\cos(\beta\rho), \qquad S(\rho)=\tan(\beta\rho), \qquad \alpha=\frac{4}{D-1}, \qquad \beta=\sqrt{\frac{(D-1)\Lambda}{2(D-2)}},
\label{alphabeta}
\end{equation}
${\phi_i\in[0,2\pi)}$, $C_i$ are the corresponding conicity parameters, and the constants $p_i$ satisfy
\begin{equation}
\sum_{i=0}^{D-2}\,p_i=1, \qquad \sum_{i=0}^{D-2}\,p_i^{\,2}=1.
\label{pis}
\end{equation}
For ${D=4}$ this reduces to the Linet--Tian solution (\ref{LinetTianmetric}), (\ref{PQdefs}) with the identification of parameters
\begin{equation}
p_0=\frac{2\sigma}{\Sigma}, \qquad p_1=\frac{2\sigma(2\sigma-1)}{\Sigma}, \qquad   p_2=\frac{1-2\sigma}{\Sigma} .
\label{identif}
\end{equation}
The ${\Lambda<0}$ counterpart of the metric (\ref{HDG})--(\ref{pis}) has been recently presented in [8,9].

\ack
We acknowledge financial support from the grants GA\v{C}R~202/08/0187, GA\v{C}R~202/09/0772, and the Project No. LC06014 of the Czech Ministry of Education.

\end{document}